\documentclass[paper]{JHEP3} 

\JHEPspecialurl{http://jhep.sissa.it/JOURNAL/JHEP3.tar.gz}

\usepackage{epsfig,multicol}

\def\bea{\arraycolsep .1em \begin{eqnarray}}
\def\eea{\end{eqnarray}}
\newcommand{\ar}{\arrowvert}
\newcommand{\ra}{\rangle}
\newcommand{\la}{\langle}

\newcommand{\cd}{\! \cdot \!}
\newcommand{\be}{\begin{equation}}
\newcommand{\ee}{\end{equation}}
\newcommand{\ba}{\begin{eqnarray}}
\newcommand{\ea}{\end{eqnarray}}
\newcommand{\pv}{{\bf p}}

\def\lsim{\mathrel{\lower4pt\hbox{$\sim$}}\hskip-12.5pt\raise1.6pt\hbox{$<$}\;}
\def\gsim{\mathrel{\lower4pt\hbox{$\sim$}}
\hskip-12.5pt\raise1.6pt\hbox{$>$}\;}

\def\s0#1#2{\mbox{\small{$ \frac{#1}{#2} $}}}
\def\0#1#2{\frac{#1}{#2}}

\title{Shear Viscosity in a CFL Quark Star}

\author{Cristina Manuel \\
{\it Instituto de F\'{\i}sica Corpuscular, C.S.I.C.-Universitat de
Val\`encia
\\ Edificio de Institutos de Paterna, Apt 2085, 46071
Val\`encia, Spain} \\ E-mail: \email{cristina@ific.uv.es} }
\author{Antonio Dobado and Felipe J. Llanes-Estrada\footnote{On leave at SLAC} \\
{\it Departamento de F\'{\i}sica Te\'orica I,  Universidad
Complutense, 28040 Madrid, Spain}\\
E-mail: \email{dobado@fis.ucm.es}, \email{fllanes@fis.ucm.es}}

\date{today}

\abstract{
We compute the mean free path and shear viscosity in the color-flavor
locked (CFL) phase of dense quark matter at low temperature $T$, when the
contributions of mesons, quarks and gluons to the transport coefficients
are Boltzmann suppressed. CFL quark matter displays superfluid properties,
and transport phenomena in such cold
regime are dominated by phonon-phonon scattering. We study superfluid phonons
within thermal field theory and compute the mean free path
associated to their most relevant collision processes.
Small-angle processes turn out to be more
efficient in slowing transport phenomena in the CFL matter,
while the mean free path relevant for the shear viscosity is less
sensitive to collinear scattering due to the presence of zero modes in the
Boltzmann equation.
In analogy with superfluid He$^4$, we find the same $T$ power law
 for the superfluid phonon damping rate and  mean free path.
Our results are relevant for the study of rotational properties of
compact stars, and correct wrong estimates existing
 in the literature.}

\received{ }

\begin{document}


\section{Introduction}

It has been known  for  long time that
cold dense quark matter should exhibit the phenomenon of color
superconductivity, at least at asymptotic large baryonic densities
\cite{Bailin}. Only recently this interesting
scenario has been considered thoroughly, both with the hope of deepening
our understanding of QCD, and for its  applications in
astrophysics (see \cite{Rajagopal:2000wf,Alford} for  reviews and
references). These recent studies realized that the fermionic gap
$\Delta$ might be as large as 100 MeV, for quark chemical
potentials $\mu$ of the order of 500 MeV.
Thus the phenomenon
should lead to clear distinct features at the macroscopic level.
Much efforts are now being devoted to finding signatures of color
superconducting quark matter in any of its different possible
phases. There are different superconducting phases, according to
the number of quark flavors that participate in the diquark
condensation. In this article we will only be concerned about the
color-flavor locked (CFL) phase \cite{Alford:1999mk}, which occurs
in the presence of three light quark flavors.

An important set of signatures of color superconductivity might be
found in phenomena associated to stellar vibration and/or
rotation. The existence of r(otational)-mode instabilities in all
relativistic rotating stars \cite{Andersson:1997xt} seems to be
incompatible with the existence of millisecond pulsars, unless the
instabilities are suppressed by sufficiently large viscosities.
Thus, the values of the viscosities can be used to rule out models
for millisecond pulsars.  While the common
belief is that pulsars are neutron stars, it cannot be discarded
that they are quark or hybrid stars.
 Madsen did point out that millisecond pulsars could be
made of strange quark matter \cite{Madsen:sx},
characterized by large shear and bulk
viscosities, but he ruled out  \cite{Madsen:1999ci}
 quark matter in  the CFL phase. This last conclusion, however, was
based on a wrong estimate, as it was assumed
that the main contribution to the viscosities was due to
gapped quarks, being exponentially suppressed as
$\exp[-2\Delta/T]$, for values of the temperature $T$ of the order of
0.1 MeV and below.

Transport properties in the CFL phase of QCD are not dominated by
the quarks. In this phase the diquark condensate breaks
spontaneously the baryon symmetry $U_B(1)$, and CFL quark
matter is a superfluid. Chiral symmetry is also spontaneously
broken. Associated to those breakings, there are Goldstone bosons,
which are light degrees of freedom.
 In addition, there is an unbroken $U(1)$ subgroup whose gauge boson,
 a combination of
the photon and one gluon, is massless at zero
temperature and can be viewed as the in-medium photon.  A CFL quark star
is electrically neutral \cite{Rajagopal:2000ff}, both for the real and
in-medium electromagnetism, but at finite temperature
one may also expect to find electrons. All
the above mentioned particles are light and their
contribution to the transport coefficients in this phase is bigger
than that of the gapped quarks. Let us mention  that we are considering
a CFL quark star after the deleptonization era,
so that all the neutrinos have escaped from the star. At higher temperatures,
the contribution of neutrinos to the transport coefficients
could be estimated from their mean free path, as
computed in refs. \cite{Reddy:2002xc,Jaikumar:2002vg}, but we will
not consider such a temperature regime here.

Chiral symmetry
is not an exact symmetry of QCD. Therefore, the
associated (pseudo) Goldstone bosons (the pions, kaons, and etas)
are massive \cite{Casalbuoni:1999wu,Son:2000cm}. At asymptotic
large densities, meson condensation may occur
\cite{Bedaque:2001je,Kaplan:2001qk}. At more moderate densities,
instanton effects become relevant, and modify the meson mass
pattern \cite{Manuel:2000wm,Schafer:2002ty}. There are still some
uncertainties in the instanton contribution to the meson masses,
however  some primary computations  indicate that all
meson masses $m_{\pi}$ are in the range of tens of MeV, or
even larger \cite{Schafer:2002ty}. One can safely state then that
at very low temperatures, $T \ll 1$ MeV, as for the estimates of
ref.  \cite{Madsen:1999ci}, the contribution of the mesons to the
transport coefficients is   also very suppressed  by the Boltzmann factor
$\exp[- m_{\pi}/T]$.

In the regime $T \ll \Delta,
m_{\pi}$, light particles dominate transport.
This was already pointed out in refs.
\cite{Shovkovy:2002kv,Shovkovy:2002sg}, where estimates for the
thermal and electrical conductivities were given.
The contribution of the in-medium electromagnetism to the shear
viscosity can be extracted from that of a QED plasma, with minor
modifications, as we carefully explain in  Sect.~\ref{QEDviscos}, but it
is small.
We report here a new computation of the contribution to the shear
viscosity  of the only truly massless Goldstone boson of the CFL phase,
the superfluid phonon. Its self-interactions
have been derived in an elegant way by Son \cite{Son:2002zn}.

Shear viscosity describes the relaxation of the momentum
components perpendicular  to the direction of transport,
and it is  usually dominated  by large-angle
collisions. In this manuscript we compute the mean free path
associated to both small and large angle collisions.
The differential
cross section of binary collisions mediated by  phonon exchange
is divergent for small-angle collisions. This is the typical
Coulomb-Rutherford collinear divergence induced by massless exchange.
In an ordinary scalar theory such a
divergence does not appear, as a thermal mass is  generated
even if the boson is massless in vacuum.
But  the phonon remains massless at finite
temperature, as thermal effects do not represent an explicit
breaking of baryon symmetry. The divergence is  regulated
by the finite width of the phonon, or more precisely,
by Landau damping, a process only occurring in a thermal bath.
After regularization, we find that small-angle
processes have a shorter mean free path than large-angle ones.
This suggests that they
might be more relevant for transport,  as a large-angle collision
can always be achieved by the addition of many small-angle ones. To compute the
shear viscosity we solve the Boltzmann equation, linearizing it
for small departures from equilibrium. Then one finds that
a zero mode occurs in the collision operator, which suppresses
severely the contribution of exact collinear processes. Then  one needs to
extend the integration  over phase space beyond the strictly collinear
limit. The full analysis is rather technical, and requires numerical
evaluation of all processes. We also find that the phonon of the
CFL and He$^4$ superfluids share many properties, and have the same
power laws for their damping and mean free path, suggesting a sort of
universal behavior.

This article is structured as follows. We review the superfluid
hydrodynamic equations and the phonon effective Lagrangian in
Sec.~\ref{lagsection}. In Sec.~\ref{selfsection}
 we compute the phonon self-energy at one-loop.
From the imaginary part of the on-shell
energy we evaluate the phonon damping rate. In Sec.~\ref{mfpsection}
we present the mean free path of splitting processes and
binary collisions. The computation of the shear viscosity is done
in Sec.~\ref{trans+shear}, and in Sec.~\ref{QEDviscos} we explain
why the contribution of the in-medium electromagnetism is
negligible at low temperatures. We comment in Sec.~\ref{corrections}
on possible corrections to our results, and present conclusions
in Sec.~\ref{discussion}.
 We leave for the
Appendices~\ref{montecarlo} and \ref{nume-shear} the technical
and numerical details of several computations.

\section{The CFL superfluid and the phonon low-energy effective theory.}
\label{lagsection}

The low energy effective theory for the only truly Goldstone boson
of the CFL phase can be constructed from the equation of state
(EOS) of normal quark matter \cite{Son:2002zn}. For three massless
quark flavors the EOS is
\be \label{freeEOS} P (\mu)
=\frac{3}{4\pi^2} \mu^4 \ ,
\ee
where $\mu$ is the quark chemical potential.
From Eq. (\ref{freeEOS}) Son obtained the effective Lagrangian
for the superfluid phonon
\be
\label{L-BGB-0} {\cal L}_{\rm eff}  = \frac{3}{4 \pi^2}
\left[ (\partial_0 \varphi - \mu)^2 - (\partial_i \varphi)^2
\right]^2 \ .
 \ee

There is an interesting  interpretation of the equations of motion
associated to $\varphi$, as they can be re-written as the hydrodynamical
conservation law of a current representing baryon number flow,
\be
\label{S-hy-1}
 \partial_\nu (n_0 u^\nu) = 0 \ ,
\ee
 where $n_0 =\frac{dP}{d \mu} |_{\mu =\mu_0}$ is interpreted
as the baryon density \cite{Son:2002zn}.
The superfluid velocity $u^\nu$ is proportional to the gradient of the
condensate phase, as in Landau's model of superfluidity
\cite{landaufluids},
\be \label{svelocity}
u_\nu = - \frac{D_\nu \varphi}{\mu_0} \ ,
\ee
where $D_\nu \varphi \equiv \partial_\nu \varphi - (\mu,{\bf 0})$,
and $\mu_0 = (D_\nu \varphi D^\nu \varphi)^{1/2}$.
The  energy momentum tensor can also be written in terms of the velocity
defined in Eq.~(\ref{svelocity}) and Noether's energy-density $\epsilon$,
 \be
\label{S-hy-2}
T^{\nu \rho} = (\epsilon + P) u^\nu u^\rho - g^{\nu \rho} P \ .
 \ee
It is conserved and traceless
\be \label{Tconserved}
\partial_\nu T^{\nu \rho} = 0\ ,\qquad  T^\nu_\nu=0 \ .
\ee

Eqs.~(\ref{S-hy-1}) and (\ref{S-hy-2}) are the hydrodynamic
equations for the relativistic superfluid  \cite{Son:2002zn}.
They need modifications at  finite temperature $T$, as
phonons are thermally excited and conform a different fluid in the
system \cite{khalatnikov,son2}.
 The well-known two-fluid model is necessary, featuring
a superfluid that flows perfectly and shows no dissipation, and a normal
fluid where  dissipative  processes are possible. At low $T$,
these  will be controlled by the
thermal properties of the phonons composing the normal fluid and
their scattering rates, that we compute in the following sections.
To proceed with those calculations we first rescale the phonon field
\be
\phi = \frac{3 \mu}{\pi} \, \varphi
\ee
to normalize the kinetic term in accordance with the LSZ formula.
Then the Lagrangian for the rescaled field reads
\be \label{L-BGB}
 {\cal L}_{\rm eff}  = \frac 12 (\partial_0 \phi)^2 -
\frac{v^2}{2} (\partial_i \phi)^2 - \frac{\pi}{9 \mu^2}
\partial_0 \phi (\partial_\mu \phi  \partial^\mu \phi) +
\frac{\pi^2}{108 \mu^4}(\partial_\mu \phi  \partial^\mu
\phi)^4 \ ,
\ee
where $v= 1/\sqrt{3}$ is the phonon velocity. We have neglected above
a total time derivative term, which is irrelevant for all computations to
follow. The Lagrangian (\ref{L-BGB}) is
not renormalizable, and only valid up to scales of the order
$2 \Delta$, the energy necessary to break a diquark condensate.
Eq. (\ref{L-BGB}) only contains the lowest derivative couplings,
valid for low energy scales. Zarembo \cite{Zarembo:2000pj}  has shown that to all orders in
$\frac{\Delta}{\mu}$, the interaction terms are  suppressed by powers of
$\mu$. Thus, for the temperature regime we will consider here $T \ll \Delta$,
those would only give very tiny corrections.

Also of interest to us is  Zarembo's observation \cite{Zarembo:2000pj} that the phonon
dispersion relation at  low momenta, beyond Son's theory,  becomes
\begin{equation}
\label{Zar-dr}
\omega_s(k) = \frac{1}{\sqrt{3}} k \left[
1 - \frac{11}{540} \frac{k^2}{\Delta^2} + {\cal O}\left( \frac{k^4}{\Delta^4}\right)
\right] \ .
\end{equation}
As $k$ is increased, the phonons move slower, with the tendency of
suppressing collinear splitting (a phonon cannot decay into two phonons
of larger joint energy). We will come back to this point in the following
sections.

\section{One-loop phonon self-energy  } \label{selfsection}

In this Section we compute  the phonon self-energy, and
evaluate its damping rate at one-loop.
We use the Imaginary Time Formalism (ITF) to perform
 the calculation. Feynman rules  are
easily deduced from the Lagrangian (\ref{L-BGB}).

The phonon propagator in the ITF  with momentum
$K= (k_0, {\bf k}) =  ( i \omega_n , {\bf k})$, reads
 \be
S(K) \equiv \frac{1}{\omega_n^2 + E^2_k}  \ ,
\ee
 where $\omega_n = 2 \pi nT$, with $n \in \cal Z$,
is a bosonic Matsubara frequency,
 and
 $E_k = v k$, where $k= |{\bf k}|$.

\begin{figure}
\begin{center}
\hbox{\psfig{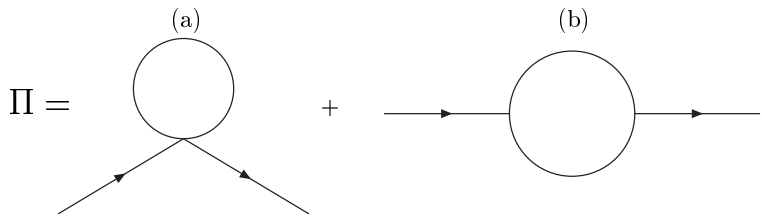}} \caption{One-loop
contributions to the phonon self-energy.
 \label{feyn2}}
\end{center}
\end{figure}

There are two different diagrams that contribute to
the one-loop self-energy, see Fig. \ref{feyn2}.
For external momentum $P = (p_0, {\bf p}) = (i \omega, {\bf p})$,
these are
\be
\Pi^{(a)} (P)   =   \frac{\pi^2}{27 \mu^4}
 \int \frac{d^d K}{(2 \pi)^d} \, \left( 2 (K \cdot P)^2 + P^2 K^2\right) S(K) \ ,
 \ee
 and
 \be
 \Pi^{(b)} (P)   =   \frac{4 \pi^2}{81 \mu^4}
 \int \frac{d^d K}{(2 \pi)^d} \, \Big( F(P,K) S(K) S(P-K)
\Big) \ ,
\ee
respectively, where
\be
\label{cub-vertex}
 F(P,K) = \left(p_0 (2
K\cdot P - K^2) + k_0 (P^2 - 2 K\cdot P) \right)^2 \ .
\ee

 Above we have used the notational convention of the ITF
\be
 \int \frac{d^d K}{(2 \pi)^d} \equiv T \sum_{n=-\infty}^{n = \infty}
 \int
\frac{d^d k}{(2 \pi)^3} \ , \ee
where the sum is over the
Matsubara frequencies. We use dimensional regularization to deal
with the ultraviolet (UV) divergences of the $T=0$ part of
the diagrams, so that $d=3 -2 \epsilon$. Since the thermal part of
the diagrams is UV finite, we will analytically continue back to
$d=3$ for their evaluation.

The superfluid phonon is a Goldstone boson, and  since thermal effects
do not represent an explicit breaking of the $U_B(1)$ symmetry, its
self-energy should vanish at $P=0$. It is actually easy to check that
this holds at one-loop level
\be
 \Pi^{(a)} (P=0) =  \Pi^{(b)} (P=0) = 0 \ ,
\ee
so no thermal mass is generated. This property of the self-energy
should hold to all orders in perturbation theory.

After performing the sum over Matsubara frequencies we find that the first
diagram only corrects the phonon velocity by a term proportional
to  $(T/\mu)^4 \ll 1$.

We proceed with the evaluation of the sum of Matsubara frequencies
for the second diagram. We find
\be \label{impart}
 \Pi^{(b)} (P)   =  - \frac{ 4\pi^2}{81 \mu^4} \sum_{s_1, s_2 = \pm}
 \int \frac{d^d k}{(2 \pi)^d} \,  F(P,K) \Big |_{k_0 = s_1 E_1}
\Big(\frac{s_1 s_2}{4 E_1 E_2} \frac{1 + f(s_1 E_1) + f(s_2 E_2)}
{i \omega - s_1 E_1 -s_2 E_2} \Big) \ . \ee
 where $E_1 = v k$ and
$E_2 = v |{\bf p - k}|$, and
\be
 f (x) \equiv \frac{1}{e^{x/T} -1}
\ . \ee

After analytical continuation to Minkowski space with retarded boundary
conditions, $i\omega \rightarrow p_0 + i \epsilon$, one notes that this
diagram has both real and imaginary parts, the last one being related
to the damping of the phonon.

The real part of the diagram gives contributions that behave as
follows for low momenta. For $p_0 = p$,  the thermal corrections
are proportional to $(T/\mu)^4 p^2$, while the $T=0$ corrections
go as $p^6/\mu^4 \ln{(M^2/p^2)}$, where $M$ is a renormalization
scale.  As for the
previous diagram, these corrections are very much suppressed as
compared to the tree
level physics for
$p \ll T \ll \mu$,  and  we will neglect them.

Although we use the full imaginary part of the
self-energy in our computer work, numerically evaluating Eq.~(\ref{impart}),
we can display an exact analytical limit for illustration.
For low external momentum, $p_0, p \ll T$, we keep only quadratic
terms in  the external momentum in  $F$, so that
\bea
 {\rm Im} \,\Pi (p_0,{\bf p}) & \approx &  \frac{4 \pi^3}{81 \mu^4}
\int \frac{d^d  k}{(2 \pi)^d} \, \frac{E_1 ({\bf k \cdot p})^2}{
E_2 }
\Big \{\left(1 + f_1 +f_2 \right)
   \Big( \delta(p_0  - E_1 - E_2) \\
 & - &  \delta(p_0 + E_1
+ E_2 ) \Big)    -  \left( f_1 - f_2
\right)\Big( \delta(p_0 -E_1 + E_2) - \delta(p_0  + E_1
- E_2) \Big) \Big\}
 \ , \nonumber
\eea
where $f_i \equiv f(E_i)$.
For $p_0, p \rightarrow 0$ it is easy to check that the last two
delta functions provide the leading order behavior to the integral,
so we neglect the first two deltas.
For the last two delta functions, we perform the following approximations
\be
E_1 - E_2 \sim v\, {\bf p \cdot {\hat k}}  \ , \qquad f_1 - f_2 \sim
 v\, {\bf p \cdot {\hat k}}  \frac{ d f_1}{d E_1} \ ,
\ee
where ${\bf {\hat k}} \equiv {\bf k}/k$. Thus, one finds for $d=3$

\be
 {\rm Im}\, \Pi (p_0,{\bf p})  \approx  - \frac{2 \pi^3}{81 \mu^4}
\int_0^\infty \frac{ k^2 dk}{2 \pi^2} \, ( k p)^2  \frac{ d
f_1}{d E_1} \int_{-1}^1 dx \,
 x^3 \Big( \delta(x - \frac{p_0}{v p}) -
\delta(x + \frac{p_0}{v p}) \Big)
 \ ,
\ee
and after evaluating the integrals, one gets
\be
\label{Ldamp-psmall}
 {\rm Im}\, \Pi (p_0,{\bf p})  \approx    \frac{8 \pi^5}{1215}
\frac{T^4}{ v^7 \mu^4} \frac{ p^3_0} { p} \, \Theta(v^2 p^2 -
p^2_0) \ ,
\ee
where $\Theta$ is the step function.
This imaginary part of the self-energy corresponds to what is
known as Landau damping. Particles  disappear or are created
through scattering in the thermal bath, and not via the processes
which occur at zero temperature.

The damping rate is defined in terms of the on-shell imaginary
part of the self-energy
\be
 \gamma(E_p) = - \frac{v^2}{2 E_p} \,
{\rm Im}\, \Pi (p_0 = E_p, \pv) \ . \ee Since we neglect the real
corrections to the self-energy, we can take $p_0 = E_p = p/\sqrt{3}$.
In the limit $p \rightarrow 0$, we find
 \be
\gamma =   \frac{ 4 \pi^5}{405 \sqrt{3}} \frac{T^4} {\mu^4} p + {\cal
O}(p^3) \ .
 \ee
Let us insist that the $T=0$ contribution to the damping rate is subleading in
$p$. The optical theorem indicates that this should be  proportional
to $\sim
p^5/\mu^4$, as the on-shell imaginary part of the self-energy is
 related to the square of the tree-level
scattering amplitude. Thus, at very low momenta, the thermal damping
dominates over the $T=0$ one (but again, we include all effects in our
numerical analysis).

Remarkably, we find that the phonons of the CFL and He$^4$ superfluids
share many properties. They both travel at the speed of sound of
the system. Their damping rates for low momenta and at low $T$ follow the
same law, going as $\sim p T^4$ \cite{Hohenberg}. Their damping
rate at $T=0$ also behaves  as $\sim p^5$ in both cases \cite{Beliaev}.
These universal properties are more easily spelled out if one writes an
effective field theory for the Goldstone mode of the non-relativistic
superfluid \cite{Andersen:2002nd}, where one finds the same sort of
 cubic and quartic derivative couplings as for the CFL superfluid.
This paralellism will be further studied in a separate publication.

\section{Phonon mean free path}\label{mfpsection}

In this Section we present explicit computations of the mean free path
associated to $1 \leftrightarrow 2$ and $2 \leftrightarrow 2$ phonon collisions.
This is a
necessary pre-analysis for the study of transport coefficients in the
system.
 In a dilute gas, viscous and other transport coefficients are
  proportional to the mean free path, and hence inversely
proportional to the damping rate or scattering cross section between
quasiparticles. Here diluteness implies that
the mean flight time between collisions is much larger than the duration
of the collision, and in addition Boltzmann's molecular chaos
hypothesis (no correlation between successive collisions) must hold.

\begin{figure}
\begin{center}
\hbox{\psfig{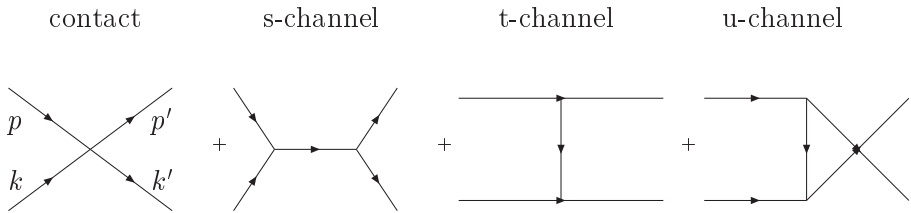}} \caption{Tree level
Feynman diagrams contributing to $2\leftrightarrow 2$ phonon
scattering processes.}
 \label{feyndiags}
\end{center}
\end{figure}

\subsection{Collinear splitting processes} \label{coll-pro}

We  evaluate the mean free path associated to
splitting collinear processes of $1 \leftrightarrow 2$ particles.
These processes  are kinematically forbidden for massive
particles, but they are not for massless ones.
Because the phonon   dispersion relation is linear with Son's 
lagrangian, these processes are perfectly collinear, that is, 
the momenta of the three particles are aligned. 
Thus, they are not efficient for shear viscosity.
They could be efficient if the phonon dispersion law was changed
so that the process were not perfectly collinear, but as 
discussed in Sec. \ref{lagsection}, corrections to Son's 
lagrangian stemming from higher derivative interaction terms in 
the effective expansion (suppressed in the $T/\mu$ counting),
disallow non-collinear splitting as shown 
by Zarembo \cite{Zarembo:2000pj}.

The inverse of the  mean collision time for these processes is obtained averaging the
one-loop damping rate computed in Sec.~\ref{selfsection}. Thus
\be
 \label{avgdamping}
 \Gamma_{1 \rightarrow 2}  = \frac{1}{n} \int
\frac{d^3  p}{(2 \pi)^3}\, \gamma(E_p) f (E_p)
 \ ,
\ee
where
\be
 n=  \int \frac{d^3  p}{(2 \pi)^3}\, f (E_p) = \frac{T^3}{4 \pi^2 v^3}\zeta (3)\ ,
 \ee
and  $\zeta (x)$ is the Riemann $\zeta$
function with $\zeta(3)=1.202...$.

The mean free path is given by $\lambda=\frac{v}{\Gamma}$.
From  simple dimensional analysis, it can be inferred that the mean free
path for these processes increases at low $T$ as $\mu^4/T^5$.
We have numerically computed  $\Gamma$ and
the mean free path with the exact one-loop self-energy,
checking this power law for low $T$ (see Fig.~\ref{meanfree}).
This may fail at higher $T$
as perturbative corrections to ${\rm Re}\  \Pi$ turn relevant,
but that temperature regime is not considered in this article.

\subsection{Two-body elastic scattering processes}

Here we compute the mean free path of binary collisions, $2
\leftrightarrow 2$
processes as depicted in Fig. \ref{feyndiags}. To evaluate  their
mean free path  one  needs the imaginary part of
 two-loop self-energy diagrams.
 We employ instead the more direct formulation
of kinetic theory, which provides the same answer. Thus, the inverse
of the mean collision time is given by

\be
\label{damping3}
 \Gamma_{2\to 2} = \frac{v^2}{2n} \int_{\bf p, k, p', k'} f_{\bf p} f_{\bf k} \left( 1 + f_{\bf p'} \right)
 \left( 1 + f_{\bf k'} \right)
  (2 \pi)^4 \delta^{(4)} (P + K - P' - K')\, \ar {\cal T} \ar^2 \ ,
\ee
where
\be
\int_{\bf p} \equiv \int \frac{d^3  p}{2 E_p (2 \pi)^3} \ ,
\ee
and $\ar {\cal T} \ar^2$ is the scattering amplitude squared.
We have also used as shorthands $f_{\pv} = f(E_p)$, etc.

Binary collisions with large or small-angle deflection behave very
differently. Let us start with the first ones. Dimensional analysis
simply suggests that these proceed with a mean free path
  of order $\mu^8/T^9$ \cite{Shovkovy:2002kv}. For the typical
values of $\mu \sim 500$ MeV and $T \sim 1$ MeV, the mean free path is
then larger than $10^5$ km, and exceeds the typical radius of a compact
star, $R \sim 10$ km. This means that these processes are totally
irrelevant for transport phenomena inside the star.

On the other hand, let us consider the $t$-channel scattering in a
binary collision, see Fig.~\ref{feyndiags} (the $u$-channel has
the same behavior). It is easy to check that the associated
 scattering matrix diverges for small-angle collisions.
 In a massless scalar theory  this divergence would be
cured through the generation of a thermal mass in the medium,
 which corrects the
low momenta behavior of the propagators. This cannot happen
for the phonons defined in Sec.~\ref{lagsection}, as these particles
remain massless at finite $T$.
However, the phonon width  regulates the divergence, as we
detail below.

Let  $Q = (\omega, {\bf q})$  be the momentum transfer, $Q = P -
P'$. Then the scattering matrix in the $t$-channel reads
 \be
\label{t-amplitude}
\ar {\cal T}\ar^2 = \left(\frac{4 \pi^2}{81 \mu^2}\right)^2
\frac{F(P,Q) F(K, -Q) }{\left(\omega^2 - v^2 q^2\right)^2 +
\left({\rm Im}\, \Pi(\omega, {\bf q})\right)^2}
 \ee
where $F$ was defined in Eq.~(\ref{cub-vertex}). We have only included
the imaginary part of the phonon self-energy, as the real part only
gives subleading corrections as already discussed.
Were ${\rm Im}\, \Pi = 0$, the scattering matrix would diverge when
$\omega^2 \to v^2 q^2$, as the exchanged phonon would be
on-shell.
This corresponds to a collision where the momenta of the scattered
particles is only slightly deflected by the collision.
This divergence is regulated by including the damping of the exchanged
phonon, which effectively amounts to a resummation of a certain class
of diagrams. Because the particles in the thermal bath have typical
energies of order $T$, the momentum transfer in the collision will behave
typically  as  $q \ll T$. So for a leading-order
estimate of the mean free path, it would suffice to include the
value given  in Eq.~(\ref{Ldamp-psmall}),
although for numerical computations we can
 calculate with the full self-energy, and check our approximations.

\begin{figure}
\begin{center}
\hbox{\psfig{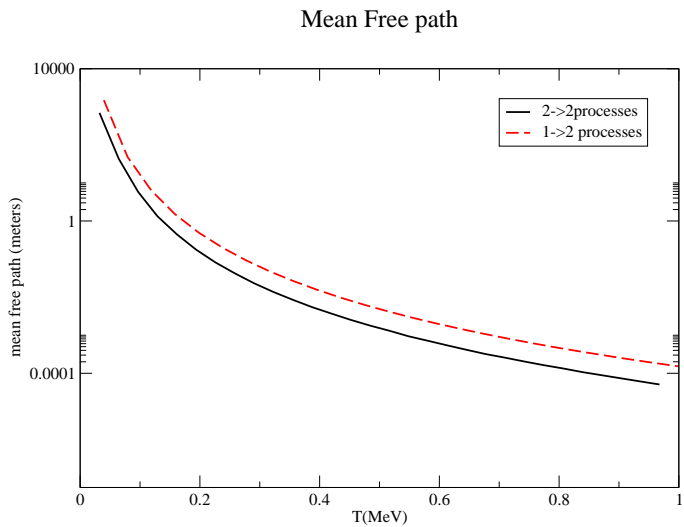}}\vspace*{1cm}
\caption{Mean free path below $T=1$ MeV, scaling  as $\mu^4/T^5$.
 At very low $T$
it exceeds the typical radius of a compact star. A hydrodynamical
description in this regime is not meaningful anymore.}
\label{meanfree}
\end{center}
\end{figure}

After regularization, the $2\to 2$ damping rate becomes
finite, and dimensional analysis of the mean free path concurs with
the full calculation on a $\mu^4/T^5$ power law (see Appendix~\ref{montecarlo}).
 The numerical estimate
(see Fig.~\ref{meanfree}) teaches us that the mean free path
of this process is only slightly smaller than that of the splitting
collinear processes.

We have excluded from the analysis the $s$-channel, as it gives a
subleading contribution to the mean free path.  As opposed to the
$t$ and $u$ channels, where the momenta transferred can be arbitrarily
small, here it is of order $T$. Thus the regulated divergence in the
$s$-channel has much less phase space in the integration region of
Eq.~(\ref{damping3}), and can be safely put aside.

Again, we have found the same $T$ dependence for the phonon
mean free path as in superfluid Helium below 0.6 K
\cite{maris}. Initially, Landau and Khalatnikov assumed that the
viscosity mean free path was governed by four phonon processes,
scaling as $~ T^{-9}$. Experimental values of the viscosity showed
that this was wrong, and that the mean free path was dominated by
small-angle collisions, behaving instead as $\sim T^{-5}$, as
in the CFL superfluid.

\section{Shear viscosity from phonon collisions} \label{trans+shear}

In this Section we compute the shear viscosity in the phonon fluid
employing  kinetic theory. The shear viscosity $\eta$ enters as a
dissipative term in the energy-momentum tensor as follows
\be
\tau^{ij} = - \eta \left( \partial_i V_j -\partial_j V_i -\frac 23
\nabla \cdot {\bf V}  \right) \ ,
\ee
 where ${\bf V}$ is the three
dimensional velocity of the phonon fluid in the frame where the
superfluid component is  at rest. The phonon four velocity is given by
$U^\mu = (\gamma, -\gamma {\bf V})$, with $\gamma^{-1} = \sqrt{1 - V^2}$.

The phonon distribution function $f$, under the hypothesis of molecular
chaos, obeys the transport equation
\be
 \label{transport}
 \frac{df_{\pv}}{dt} = C[f_{\pv}] \ .
\ee

In the frame where the superfluid is locally at rest, the
Liouville operator appearing in Eq.~(\ref{transport}) (and thus
the equilibrium solutions) depends on the superfluid velocity
Eq.~(\ref{svelocity}). Because we are interested in the shear
viscosity of the phonon fluid, and we will linearize the transport
equation in gradients of ${\bf V}$, we will neglect  the
superfluid velocity\ in all the developments to follow. We note,
however, that this approximation would not be valid for the
computation of other transport coefficients.

The collision term refers to the binary collisions and
splitting/joining
processes described in  Sec.~\ref{mfpsection}.
In the shear viscosity calculation, we neglect the last one for the reasons
explained in Sec.~\ref{coll-pro}.
 The binary collision term is given by:
 \be \label{collterm}
 C[f_{\pv}]
= \frac 12 \int_{\bf k, p', k'}
  (2 \pi)^4 \delta^{(4)} (P + K - P' - K')\, \ar {\cal T} \ar^2
\frac{1}{2E_p} D \ ,
\ee
with
\be
 D =  f_{\bf p'}f_{\bf k'}(1+ f_{\pv})(1+
f_{\bf k}) -f_{\pv}f_{\bf k} (1+ f_{\bf p'})(1+ f_{\bf k'}) \ .
 \ee

 As usual, the collision term
vanishes for the equilibrium distribution $f^{\rm eq.}$, which is
given by
\be  \label{BErest}
f^{\rm eq.}_\pv =\frac{1}{e^{P_\mu U^\mu/T}-1} \ ,
\ee
where $P_\mu = (E_p, \pv)$.

For the computation of  transport
coefficients we have to consider small departures from equilibrium
so that the distribution function can be written as
\be \label{outofeq}
f= f^{\rm eq.} + \delta f \ .
\ee
Accordingly, $D$ can be written as $D= D^{\rm eq.} +\delta D$.

To solve the transport equation we employ the Enskog expansion. To first
order we consider the perturbation only on the collision term and
linearize it, whereas in the advective term we take the equilibrium
distribution function.
Since the equilibrium part $f^{\rm eq.}$ annihilates the collision
term, we substitute $D$ in Eq.~(\ref{collterm}) by
$\delta D$, that can be written as
 \be
  \delta D=f^{\rm eq.}_{\pv}f^{\rm eq.}_{\bf k}f^{\rm eq.}_{\bf p'}
f^{\rm eq.}_{\bf k'}
e^{(E_p+E_k)/T}\Delta
[\frac{\delta f}{f^{\rm eq.}} (1-e^{- E/T})] \ ,
 \ee
  where the symbol $\Delta g$ is a shorthand for:
\be
\Delta g \equiv g({\bf p'})+g({\bf k'})-g(\pv)-g({\bf k})\ .
\ee

From the kinetic point of view, the dissipative part of the
 spatial momentum-stress tensor can
be written as:
\be
\tau_{ij}= \int \frac{d^3p}{(2 \pi)^3 E_p} p_i p_j \,\delta f \ .
\ee

To simplify the computations, we assume a velocity profile of the
form ${\bf V} = (V_x (y), 0, 0)$. We then assume that
the driving shear departures from the equilibrium distribution function
can be parameterized as
\be \label{param1}
\delta f = -\frac{f^{\rm eq.}}{T}  g(p) p_x p_y
\frac{ d V_x}{d y} \ .
\ee

Solving the linearized transport equation for $g(p)$
 would yield the shear  viscosity, which is  expressed as
\be
 \eta=\frac{4 \pi }{15 T v}\int
\frac{dp}{(2\pi)^3} p^5 g(p) f^{\rm eq.}  \ .
\ee

This expression can be viewed as a scalar product
\be \label{viscovar1}
\eta= \la \chi \ar \Phi \ra
\ee
with
\be
\chi = p_i p_j
\ee
and $\Phi$ a solution to the Boltzmann  equation
that can be written as
\be
\ar \chi \ra = {C} \ar \Phi \ra
\ee
with $C$ the linearized collision operator.
Equivalent to Eq. (\ref{viscovar1}) is then
\be
\eta= \la \Phi \ar C \ar \Phi \ra
\ee
that squared, and substituting (\ref{viscovar1}), yields
\be \label{viscovar2}
\eta = \frac{\ar \la\Phi \ar \chi \ra \ar^2}{\la \Phi \ar
C \ar \Phi \ra }\ .
\ee
In this expression, $\Phi$ is the exact solution of the Boltzmann
equation. But it is also amenable to a variational treatment considering a
set of test functions $\bar{\Phi}$, where the exact solution $\Phi$ would
make (\ref{viscovar2}) reach an extremum (as can be seen by substituting
in it $\bar{\Phi} = \Phi + \epsilon$ and observing that the linear terms in
$\epsilon$ cancel if $\Phi$ solves the transport equation).

In a rotationally invariant way, we can take the scalar product necessary
in these expressions as
\be
\cdot  = v^2\int \frac{d{\bf p}}{E} \frac{f_0}{T} \frac{1}{10} \sum_{ij}
\ee
and we take as trial function
\ba
\Phi= g_{ij} (p)= g(p) P_{ij}=
g(p) (p_ip_j - \frac 13 \delta_{ij}p^2)
\\
\nonumber
g(p)= \frac{p^\nu}{1-e^{-E_p/T}}
\ea
with $\nu$ a variational parameter. This family is general enough to
allow treatment of a  polynomial family such as used in
\cite{Dobado:2003wr}, but also includes rational functions.
For this family, Eq.~(\ref{viscovar2}) reduces to
\be \label{viscovar3}
\eta[\nu]=\frac{(I[\nu])^2}{L[\nu]} \ ,
\ee
with
\be
I[\nu]=\frac{v^2}{15}\int \frac{d p}{E} p^4 \frac{f^{\rm eq.}}{T}
\frac{p^\nu}{1-e^{-E_p/T}} \ ,
\ee
and
\ba \label{COLLISIONTERM}
L[\nu]= \frac{1}{40T} \int_{\bf p, k, p', k'}
(2\pi)^4 \delta^4(P+K-P'-K') \frac{\ar
T\ar^2}{2} \\ \nonumber
e^{(E_p+E_k)/T} f^{\rm eq.}_{\bf p} f^{\rm eq.}_{\bf k}f^{\rm eq.}_{\bf p'}
f^{\rm eq.}_{\bf k'} \Delta(P_{ij}p^\nu)
\Delta(P_{ij}p^\nu) \ .
\ea
As the integrand of $L[\nu]$ is positive definite,
so is also Eq. (\ref{viscovar3}). The computation of $I[\nu]$ is easy, but
to compute $L[\nu]$  requires a multidimensional Montecarlo integration.
This is quite a technical calculation and is detailed in the Appendices,
but we will point out some subtleties here.

It is worth stressing that for a quick order of magnitude
estimate of the shear viscosity,  it is usually correct to
pick up as a trial function the case considered for $\nu = 0$.
It usually gives also the correct parametric behavior of the viscosity.
However, in this case this is not so, as some values of the parameter
$\nu$ cause a zero-mode in the collision term\footnote{We
thank the referee for calling this important point to our attention.}.
We have
identified two of these zero-modes.

Let us first see why these zero modes occur. For a general trial function,
the denominator of Eq.~(\ref{viscovar3}), and as explained at length in
the previous sections, is dominated by contributions to the integral coming
from (almost) collinear collisions due to the  divergence
in the naked $t$-channel amplitude. This divergence is however regulated
by the inclusion of Landau damping, see Eq.~(\ref{t-amplitude}).
For a perfect collinear collision,  the
conservation of energy momentum at the vertex $\delta^{(3)}({\bf
k}+{\bf q}-{\bf p}) \delta(E_k+\omega-E_p)$, where $\omega$ and ${\bf q}$
are the energy and momentum transfer,
 together with $E_p=v p$
requires that ${\bf p}$, ${\bf k}$ and ${\bf q}$ are all parallel.
Therefore all entering and outgoing boson momenta in a collinear
collision are parallel. However, in
solving the transport equation,  a special choice of trial wavefunction
may force a zero in this collinear limit and it can thus achieve a
cancellation.
Then this zero mode dominates the inversion of the collision operator $C$  when
solving the Boltzmann equation.

This property of the (linearized) collision operator stems from the term
\be
\sum_{ij} \Delta(P_{ij} p^\nu)\Delta(P_{ij} p^\nu) \ .
\ee
Now observe the cancellation for $\nu=-2$. In this case,
$P_{ij} p^{-2}= \hat{p}_i \hat{p}_j - \frac{\delta_{ij}}{3}$
depends only on the unit vector along the momentum, and
\be
 \Delta(P_{ij} p^\nu) = \hat{p}_{i}\hat{p}_{j} +
\hat{k}_{i}\hat{k}_{j}- \hat{p'}_{i}\hat{p'}_{j}
-\hat{k'}_{i}\hat{k'}_j = 0 \ .
\ee
for a perfect collinear collision.  A second zero-mode is given by the variational value $\nu=-1$. Then
\be
 \Delta(P_{ij} p^\nu) = \frac 1v \left( \frac{ }{ }
E_p \hat{p}_{i}\hat{p}_{j} +
E_k \hat{k}_{i}\hat{k}_{j}
- E_{p'} \hat{p'}_{i}\hat{p'}_{j} -E_{k'} \hat{k'}_{i}\hat{k'}_{j}
+\frac 13 \delta_{ij} (E_p +E_k -E_{p'}-E_{k'}) \right).
\ee
This also vanishes in the collinear limit upon employing the
energy-conservation equation.

The collision operator does not vanish because contributions to
the integral not in the perfect collinear limit  still
provide a contribution, and thus a near-zero mode, not an exact zero mode,
appears.

We have carried out a variational study of $\eta[\nu]$ numerically
(see Appendix \ref{nume-shear}), and we have reached to the conclusion
that $\eta$ is maximized for the value $\nu = -1$, which corresponds to
one of the (near) zero-modes explained above.

We quote here the final outcome of these analysis.
Due to the existence of the (near) zero mode with $\nu=-1$
that suppresses collinear scattering, all $2 \leftrightarrow 2$
channels are included, and the viscosity behaves as
\be
\label{final-all}
\eta= 1.3\cd 10^{-4} \cd \frac{\mu^8}{T^5} \ {\rm MeV}^3 \ .
\ee

For illustration we also give the result for low momentum transfer that
is amenable to analytical treatment  when only the $t$-channel is
included  the viscosity behaves instead as

\be
\label{final-t}
\eta \sim 4  \cd 10^{12} \cd
\frac{\mu^8}{T^5 \log{|T/\mu|}} \, {\rm MeV}^3 \ .
\ee

\section{Contribution of the in-medium electromagnetism to the
viscosity}
\label{QEDviscos}

In considering transport coefficients for the CFL quark star we should
examine the contribution of the in-medium electromagnetism,
 since the very light electron could
conceivably transport momentum efficiently.
 The effective coupling of the in-medium
photon and the electron is
${\widetilde e} = e \cos{\theta}$, where $e$ is the electromagnetic
coupling constant, and the mixing angle is \cite{Litim:2001mv}
\be
\cos{\theta} =  \frac{\sqrt{3} g}{\sqrt{3 g^2 + 4 e^2}} \ ,
\ee
CFL quark matter is a dielectric at low energies, with
dielectric constant \cite{Litim:2001mv}
\be
\widetilde\epsilon
= 1 + \frac{{2\,}}{9 \pi^2}\frac{\widetilde e^2\mu^2}{\Delta^2} \ .
\ee
In particular, this implies that the in-medium photon travels
at a speed ${\widetilde v}_{\gamma} = 1/\sqrt{\widetilde \epsilon}$,
less than the speed of light.  Reflection and refraction properties
of light on  CFL quark matter have been studied in \cite{Manuel:2001mx}.

 The value of the shear
viscosity can be borrowed from QED, by taking into account the
following replacements in the electromagnetic fine structure constant
\be
\alpha \rightarrow {\widetilde \alpha} = \frac{{\widetilde e}^2}{4 \pi
\sqrt{\widetilde \epsilon}}= \alpha \frac{\cos^2{\theta}}{\sqrt{\widetilde \epsilon}}
\ee

Transport properties in a photon-electron system in the regime $T
< m_e$, where $m_e = 0.5$ MeV is the electron mass,  have been computed in
\cite{deGroot}.
While the computation, based on electron-in medium photon 
scattering  at an arbitrary temperature can only be
done numerically, approximate analytical results give
 \be \label{shearem}
 \eta^{\rm e.m.} = \frac{20 x_\gamma}{27 x_e} \frac{ T}{
\sigma_T} \ ,
 \ee
 where $\sigma_T$ is the Thomson cross-section
\be \sigma_T = \frac 83 \pi \left (\frac{ {\widetilde
\alpha}}{m_e }\right)^2 \ ,
\ee
and $x_{\gamma,e}= n_{\gamma,e}/(n_{\gamma} + n_e)$ measures the relative
concentrations of photons/electrons. 

For an order of magnitude estimate, one can take ${\widetilde
\alpha} \sim 1/137$, as for moderates densities ${\widetilde
\epsilon}$ never becomes too large. Then, one can see that the
electromagnetic  contribution to the shear viscosity
 is negligible compared to the phonon contribution at 
temperatures of order $0.1$ MeV. Further lowering the 
temperature makes Eq. (\ref{shearem}) increase rapidly because
of the Boltzmann density factor $e^{-m/T}$ in the denominator 
containing the electron density. However, light by light 
scattering processes come to play (these can be treated with the 
Euler-Heisenberg effective lagrangian as reviewed in 
\cite{dobadoandco}) and the photon scattering cross section is 
not exponentially low.

\section{Further improvements \label{corrections}}

Let us comment on possible corrections to our results.
 First, one could better determine the
 self-interactions of the
phonons. According to ref.~\cite{Son:2002zn}, corrections to the
effective Lagrangian Eq.~(\ref{L-BGB}) can be obtained with a
better determination of the EOS of CFL quark matter. Nevertheless,
as the EOS is dominated by a term going as $\sim \mu^4$,
corrections to the phonon Lagrangian will not modify the leading
parametric dependence on $\mu$ of the three-phonon vertex, the relevant
one for transport in the superfluid, and
would only alter the numerical factor in front of this term in the
effective Lagrangian.

Second, one could better determine the collision term entering
the transport equation. We saw that Landau damping regulates
an otherwise divergent scattering matrix of binary collisions (see
Sec.~\ref{selfsection}).
Taking into account the width of the exchanged phonon
amounts to a resummation of diagrams, where multiple
splitting/joining  scatterings are treated as independent
classical events. Since the mean free path associated to the $2
\leftrightarrow 2$ and the $1 \leftrightarrow 2$ collisions are of the
same order, this approximation is not quite correct.
One should take into account the Landau-Pomeranchuk-Migdal (LPM)
effect, which would better determine the
collision term \cite{Landau:gr,Landau:um,Arnold:2002zm}. This
correction would not modify the parametric behavior of  $\eta$,
but would again affect the preceding numerical
factor.\footnote{We thank L. Yaffe for this observation.}

One should as well be concerned about possible changes in the
shear viscosity when either
higher derivative interactions  or
higher order corrections are included in
the phonon dispersion relation. As already discussed in Sec.~\ref{lagsection},
 Zarembo \cite{Zarembo:2000pj} has shown
that to all orders in $\frac{\Delta}{\mu}$, the interaction terms are
suppressed by powers of $\mu$, not $\Delta$.
Also of interest to us is  Zarembo's observation that the phonon
dispersion relation is given by Eq. (\ref{Zar-dr}).
As $k$ is increased, the phonons move slower.
The negative coefficient in the correction to the linear law above
reveals that
corrections to the dispersion relation beyond Son's theory suppress
collinear splitting (a phonon cannot decay into two phonons of larger
joint energy at $T=0$). Therefore we are safe in ignoring the $1\to 2 $
processes in our computation of the shear viscosity.

We note that for superfluid He$^4$, the phonon dispersion relation used in
ref. \cite{maris} to match the experimental value of the shear viscosity
 is of the form
\begin{equation}
\omega_s(k) = v k \left[
1 + g(k)
\right] \ .
\end{equation}
with $g(k)$ a positive function that tends to zero for $k
\rightarrow 0$. In this way, phonons with high momenta move faster
than those of low momenta  and can decay into
 two slower phonons.
Maris used the experimental value of the viscosity to gain information
about the function $g(k)$. It then turns out that
 in superfluid He$^4$  almost collinear processes
are more relevant than large-angle collisions for the shear viscosity
at very low temperature.

For astrophysical applications  an order of magnitude
estimate for the values of the viscosities is good enough to
study  properties of  compact stars.
Other transport properties would also be needed, but they will be
the subject of a different project.

\vfill

\section{Discussion} \label{discussion}

We finally summarize our main findings and conclusions.
CFL quark matter at very low $T$ behaves as a superfluid of the sort
He$^4$, rather than of the sort of He$^3$, as wrongly assumed in
the literature. Dissipative processes in such a cold regime
are dominated by phonon-phonon scattering. This analogy is reflected in
$T$-power laws for the thermal properties of phonons of
the CFL and He$^4$ superfluids.

For astrophysical applications, we should emphasize the following.
 At sufficiently low $T$ the phonon mean free path would exceed
the radius of a compact star. We can give a crude estimate of
the temperature when this will occur, simply by considering the equation
\be
 R < L \sim \mu^4 / T^5 \ .
\ee
If the quark chemical potential is of order $\mu \sim 500$ MeV,
and we consider $R \sim 10$ km, we find that for $T < 0.06$ MeV
superfluid phonons do not scatter within the star.
 Transport coefficients could then be dominated by the tiny contribution
of the in-medium electromagnetism, but
an evaluation of the photon  mean free path also shows
that for  $T \sim 0.02$ MeV it also exceeds the radius of the star
\cite{Shovkovy:2002sg}. Below that temperature,
 CFL quark matter in the star would behave as a perfect superfluid,
as a hydrodynamical description of the phonon and electron fluids
would be meaningless. The superfluid hydrodynamical equations would then
be given by Eqs.~(\ref{S-hy-1}) and (\ref{S-hy-2}),
showing then no dissipation.

In a rotating superfluid there are vortices. To study the rotational
properties of a hypothetical CFL quark star, one cannot obviate that
fact. In view of our results,  the analysis of r-mode instabilities
of a CFL quark star should then be redone, taking into account both
the temperature regime of the star, and the vortex dynamics of the
CFL phase. The bulk viscosity should also be needed,
but we leave this computation for a different project.

\acknowledgments

 We thank D. T. Son, R. Pisarski and  L. Yaffe for very useful
discussions. C.~M. thanks the I.N.T. at the University of
Washington for its hospitality and partial support during the
completion of this article.
 This work has been supported by grants FPA 2000-0956, BFM 2002-01003,
FPA2001-3031 from MCYT (Spain).

\appendix
\section{Evaluation of the  $2\to 2$ processes} \label{montecarlo}

In this Appendix  we give some details on the Montecarlo numerical evaluations of
the multiple integrals appearing in two-body elastic scattering.
Here we treat Eq.~(\ref{damping3}) at length. The integral is dominated by t-channel
scattering (the u-channel contribution is equal and amounts to a factor 2)
due to the collinear singularity.
Then for production runs it is sufficient to maintain the
$t$-channel scattering matrix $\ar{\cal T} \ar^2$ and neglect the rest
of the interaction diagrams.
The numerator of Eq.~(\ref{damping3}) has  twelve integrations.
Three, for example over ${\bf k'}$, can be immediately performed with the
help of the three-momentum $\delta$ in Eq. (\ref{damping3}).
 We choose for the nine remaining
integrals the incoming three-momenta $\bf p$, $\bf k$, and the transferred
momentum ${\bf q}= \pv-{\bf p'} = {\bf k'}-{\bf k}$.

Our freedom to choose the third axis makes
the angular integrals around $\bf q$ to be $4\pi$.
Since we approximate the integral to be dominated by  $t$-channel
exchange, the scattering matrix element $\ar {\cal T} \ar^ 2$ can only
depend on the two polar angles (through the derivative coupling in the
vertex), $\hat\pv \cd \hat{\bf q}$ and $\hat{\bf k}\cd \hat{\bf q}$ and
not on the
azimuthal angles $\phi_\pv$ and $\phi_{\bf k}$. Therefore the latter are
also trivial.

Introduction of an auxiliary variable $\omega$ (energy transfer)
allows
to write the energy $\delta$ in Eq. (\ref{damping3}) as
\be
\delta(E_ k+E_p -E_{k'}-E_{ p'}) =\int_{-\infty}^{+\infty}d\omega\, \delta(w-E_p+E_{ p'})
\delta(\omega + E_{k}-E_{ k'})\ .
\ee
Then it is easy to employ the resulting two $\delta$ functions to compute
the remaining polar angular integrals. In exchange, the integral over the
auxiliary variable remains.
With these manipulations we obtain
\be \label{fullintegral}
\Gamma(T)= \frac{1}{vT^3\pi^3\varsigma(3)2^{8}}
\int_0^\infty dq \int_{-vq}^{vq} d\omega
\int_{\frac{vq-\omega}{2v}}^{\infty} dk
\int_{\frac{vq+\omega}{2v}}^{-\infty} dp
f_{\pv} f_{\bf k} (1+f_{\bf k'}) (1+f_{\bf p'}) \ar {\cal T}\ar^2 \ ,
\ee
where $\ar {\cal T}\ar^2$ is given in Eq.~(\ref{t-amplitude}).
This is an integral in four dimensions dominated by the region around the
on-shell singularities for $\omega\simeq vq$ and $\omega \simeq -vq$. To
obtain good numerical convergence these  singularities, which are
regulated  by  Landau damping, need to be isolated. First, perform one more
change of variables to
\be
x=\omega+vq\ ; \ \ y=\omega-vq \ .
\ee
The integrals over $x$ and $y$ run from $0$ to $\infty$ and from $-\infty$
to $0$ respectively. The on-shell singularities occur now at $x=0$, $y=0$.
Therefore we introduce a low momentum cut-off on these variables,
$\lambda$, that serves to split the regions of integration in the $xy$
plane as (1) $(0,\lambda)\times(-\lambda,0)$; (2) $(0,\lambda)\times(-\infty,-\lambda)$;
(3) $(\lambda,\infty)\times (-\lambda,0)$; (4) $(\lambda,\infty)\times (-\infty,-\lambda)$.
The value of the integral should of course be independent of the
value of $\lambda$ since this merely splits the integration
region. Choosing $\lambda \ll T$ allows to evaluate most of the
integrand in regions (2) and (3) at $x=0\ , y=0$ respectively, and
therefore the $x$ ($y$) integral can be performed analytically.
Explicit numerical evaluation shows that regions (1) and (4) are
negligible with respect to the contributions from regions (2) and (3)
whenever $\lambda$ is large compared to $\frac{{\rm Im}
\Pi\ar_{y=0}}{x}$ (or the same expression reversing $x$ and $y$).
Under these conditions we need to concentrate only on regions (2)
and (3).

For region (3) the relevant $y$ integral is
\be \label{analyticalform}
 I \approx F(P,x,y=0)
\int_{-\lambda}^0 dy \frac{1}{x^2y^2+({\rm Im}\Pi(x,y=0))^2} \ .
 \ee
Since in this region $x>\lambda$, the imaginary part takes a
constant value as $y\to 0$. Therefore the integral is regulated
and yields an arctangent. All the derivative couplings and
Bose-Einstein factors are smooth at $y=0$ and have been pulled out
of the integral as a constant (still function of $x$) $F(P,x,y=0)$,
since the interval is small respect to $T$. Therefore
\be
 I=F(P,x,y=0) \approx
\frac{\pi/2}{x\, {\rm Im}\Pi(x,y=0)} \ .
\ee
 Finally the damping rate
can be given  to very good accuracy in terms of three quadratures
\ba
\label{quadratures}
\Gamma &= & \frac{1}{128\varsigma(3)
v^2T^3}\frac{\pi}{(81\mu^4)^2} \left(\bar\Gamma_2+
\bar\Gamma_3 \right) \\ \nonumber
 \Gamma_2 & \simeq &
\int_{-\infty}^{-\lambda}dy \int_{-y/2v}^\infty dk \int_0^\infty
dp  f_{\pv} f_{\bf k} (1+f_{\bf k'}) (1+f_{\bf
p'})\frac{\pi/2}{y\, {\rm Im}\Pi(x=0,y)}  \left(F(P,-Q)
F(K,Q) \right)\ar_{x=0} \\ \nonumber \Gamma_3 & \simeq & \int_{\lambda}^{\infty}dx
\int_{x/2v}^\infty dp \int_0^\infty dk  f_{\pv} f_{\bf
k} (1+f_{\bf k'}) (1+f_{\bf p'})\frac{\pi/2}{x \,{\rm Im}\Pi(x,y=0)}
\left(F(P,-Q) F(K,Q) \right)\ar_{y=0} \ .
  \nonumber
 \ea
The remaining three integrals are routinely
calculated with a computer code and yield the damping rate.
Evaluation of the shear viscosity is dealt with in the next two sections.

\section{Numerical evaluation of shear viscosity}
\label{nume-shear}

\begin{table}
\caption{\label{IRcancel} Numerical cancellation of the internal $\lambda$
dependence in $\eta(-2)$ given by eq. (\ref{viscovar3}) employing the zero
mode trial wavefunction $\nu=-2$. The temperature is $T=0.001$  MeV, the
cutoff for external momenta is $\Lambda=100 T$ and the cutoff over the
$x$, $y$ integrations is $0.6 \Lambda$, large enough to cover all relevant
phase space but small enough to avoid the Montecarlo points to be
dispersed where the Bose-Einstein factor damps the integrand. The
calculation is performed with 256000 points in each integration domain,
doubled four times for convergence, and the integral is computed ten times
for each fixed number of points. The standard deviation obtained is 1\% .
A small logarithmic drift of the result is still observed, but this is now
of the order of the uncertainty in the Montecarlo and not very relevant.
A meaningful cutoff $\lambda$ needs to be smaller than $T$ but of a
similar order of magnitude.}
\begin{center}
\begin{tabular}{|cc|}
\hline
$\lambda$ & $\eta(\nu=-2) (MeV^3)$ \\
\hline
$T/12$    & $1.09 \cd 10^{33}$\\
$T/25$    & $1.10 \cd 10^{33}$\\
$T/50$    & $1.12 \cd 10^{33}$\\
$T/100$   & $1.14 \cd 10^{33}$\\
$T/200$   & $1.16 \cd 10^{33}$\\
\hline
\end{tabular}
\end{center}
\end{table}

Our Montecarlo program, employing the subroutine Vegas
\cite{Lepage:1980dq} has been written in two versions. One perfoms three
and four dimensional integrals keeping only the t-channel
exchange contribution to the cross section as previously described, and is
useful to isolate this precisely. If other scattering processes also
contribute, this program cannot be used since two azimutal angles are now
not trivial. In this case a second version performs the
five dimensional integral over all phase space and is a generic purpose
integrator (but diluting the available computing power over a larger
region makes it less precise to capture t-channel singularities).
In table \ref{tablafinal} where we present various numerical results,
these two versions are denoted as (t-channel) and (all) respectively.

\begin{table}
\caption{\label{tablafinal} Numerical output of our
programs for shear viscosity. The program marked (t-channel) ignores
contributions from s-channel exchange and contact terms, as well as
crossed terms between them when evaluating $\ar T\ar^2$. Therefore
it evaluates only three and four-dimensional integrals as opposed to
(all) that performs five-dimensional integration.  The cutoff for
external momenta is $100 T$, the cutoff for transferred $q$, $\omega$ is
$60T$, and for the (t-channel) version, the infrared separation for the
very nearly collinear zone is at $\lambda=T/100$. All units are 
$MeV$.}
 \begin{center}
\begin{tabular}{|cr|cc|}
\hline
$T$    & Variational parameter $\nu$& $\eta$ (t-channel) & $\eta$ (all)\\
\hline
$10^{-3}$& $-3$& $9.7\cd 10^{12}$& $2.3\cd 10^{31}$\\
$10^{-3}$& $-2$& $1.1\cd 10^{33}$& $1.3\cd 10^{32}$\\
$10^{-3}$& $-1$& $1.1\cd 10^{48}$& ${\bf 5.0\cd 10^{32}}$\\
$10^{-3}$& $ 0$& $4.8\cd 10^{13}$& $1.6\cd 10^{30}$\\
$10^{-3}$& $ 1$& $2.9\cd 10^{12}$& $8.6\cd 10^{28}$\\
$10^{-3}$& $ 1$& $3.6\cd 10^{11}$& $9.9\cd 10^{27}$\\
$10^{-2}$& $-3$& $1.0\cd 10^{12}$& $2.3\cd 10^{26}$\\
$10^{-2}$& $-2$& $1.4\cd 10^{28}$& $1.3\cd 10^{27}$\\
$10^{-2}$& $-1$& $1.4\cd 10^{43}$& ${\bf 5.0\cd 10^{27}}$\\
$10^{-2}$& $ 0$& $4.8\cd 10^{12}$& $1.6\cd 10^{25}$\\
$10^{-2}$& $ 1$& $2.9\cd 10^{11}$& $8.6\cd 10^{23}$\\
$10^{-2}$& $ 2$& $3.6\cd 10^{10}$& $9.9\cd 10^{22}$\\
$10^{-1}$& $-3$& $1.1\cd 10^{11}$& $2.3\cd 10^{21}$\\
$10^{-1}$& $-2$& $2.6\cd 10^{23}$& $1.3\cd 10^{22}$\\
$10^{-1}$& $-1$& $1.9\cd 10^{38}$& ${\bf 5.0\cd 10^{22}}$\\
$10^{-1}$& $ 0$& $4.8\cd 10^{11}$& $1.6\cd 10^{20}$\\
$10^{-1}$& $ 1$& $2.9\cd 10^{10}$& $8.6\cd 10^{18}$\\
$10^{-1}$& $ 2$& $3.6\cd 10^{9}$ & $9.9\cd 10^{17}$\\
\hline
\end{tabular}
\end{center}
\end{table}
The numerical results from our Montecarlo for the shear viscosity are
given in full in table \ref{tablafinal}. The variational calculation is
somewhat complex and care needs to be exercised upon its interpretation.
In this table, the first column gives the temperature in MeV. The second
is the variational parameter $\nu$ that varies from a strong infrared
enhancement to polynomial suppression. The fourth column displays the
values obtained for the viscosity with all interactions included.
These are extracted in the third column.

 The fourth column
displays very clearly the $\frac{\mu^8}{T^5}$ behaviour
characteristic of the naive power counting. However, for a general
$\nu$, one should go to the result of the t-channel exchange.  For
example, for $\nu=0$ that corresponds to the usual zero'th order
approximation in a polynomial expansion, and at $T=0.1$ MeV,
the viscosity would behave as $\eta\simeq 0.8 \cd
\frac{\mu^4}{T}$.

But this conclusion would miss the effects of the zero mode. By studying
$\eta(\nu)$ with the t-channel calculation (again the third column, at
fixed $T$), one observes that the zero modes discussed above corresponding
to $\nu=-2$ and $\nu=-1$ yield variational maxima, especially $\nu=-1$, a
mildly dependent on the numerical grid and cutoffs. The analytical
considerations suggest a logarithmic dependence, as discussed in the
main text.

\begin{figure}
\begin{center}
\hbox{\psfig{file=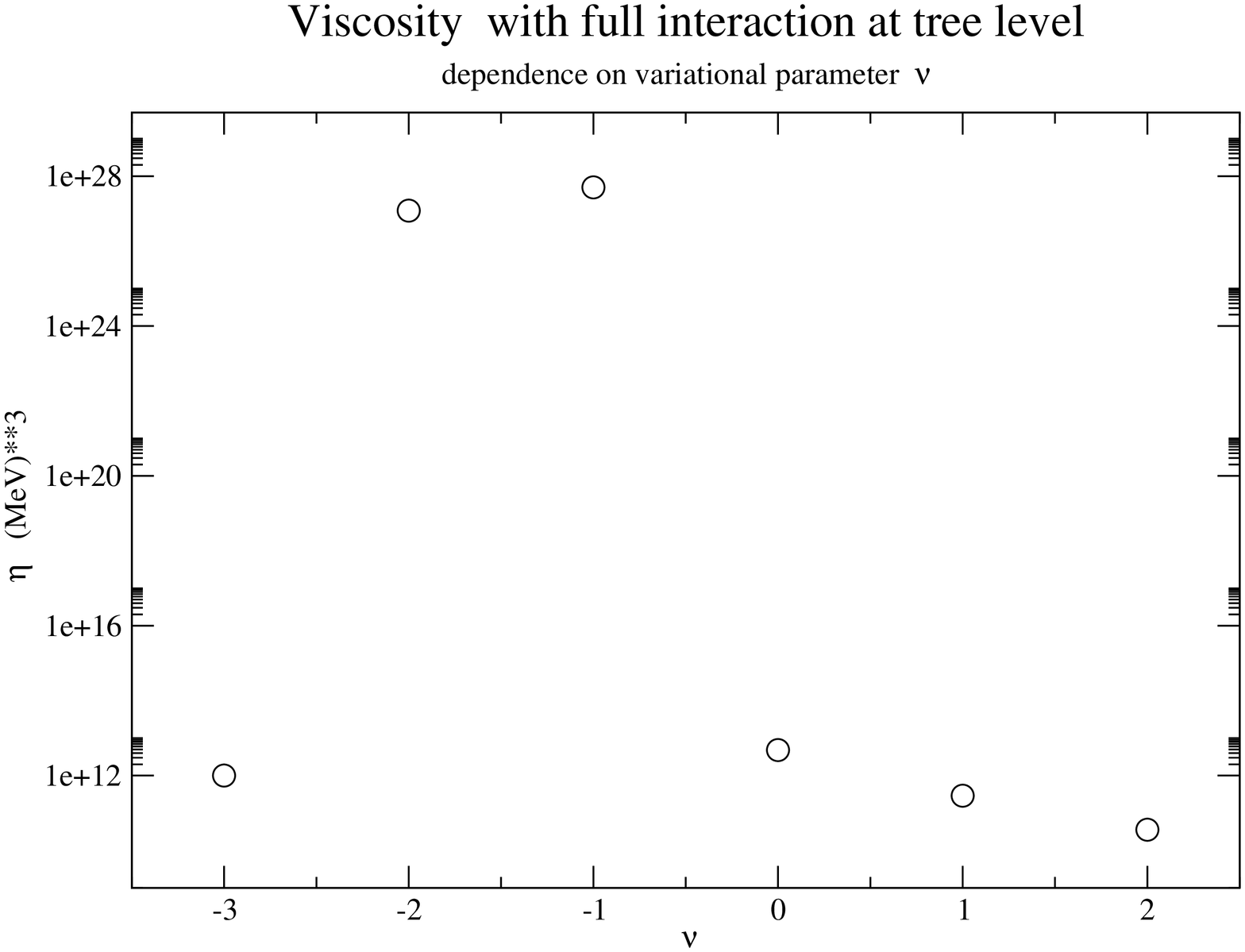,angle=-90,width=18cm}}
\caption{\label{full}
The variational evaluation including boson
exchanges in the interaction
shows a maximum corresponding to the zero modes that make the t-channel
contribution suppressed.
The maximum within the functional family $\frac{p^\nu}{1-e^{-E_p/T}}$
corresponds to $\nu=-1$. The Temperature is fixed at 0.01 $MeV$.}
\end{center}
\end{figure}

Finally a glance at
the fourth column reveals that this formula is much larger (and therefore
the integral (\ref{COLLISIONTERM}) very suppressed) than the corresponding
naive counting for this value of the variational parameter. This entry
also happens to be the maximum of the third column, finally concluding
that the viscosity must behave in this temperature range as in Eq.(\ref{final-all}).
The plot \ref{full} enlightens this discussion by showing how the
t-channel contribution is much lower than the corresponding $T^{-5}$
behaviour for $\nu=-1$, $\nu=-2$.

\end{document}